# Noctilucent Clouds Modulated by Strong 5-day Planetary Wave in 2025: Amplitudes, Phases and Altitudes Based on Ground-Based Observations and Satellite Temperature Data

Oleg S. Ugolnikov[a*], Ilya S. Yankovsky[b], Nikolay N. Pertsev[c], Vladimir I. Perminov[c], Maxim V. Klimenko[b], Ekaterina N. Tipikina[d], Alexey V. Popov[e], Andrey M. Tatarnikov[f], Sergey G. Zheltoukhov[f], Sergey A. Potanin[f], Egor O. Ugolnikov[g], Olga Yu. Golubeva[h], Andrey L. Kotikov[i], Alexey S. Sushkov[j], Egor A. Volkov[j]

[a] *Space Research Institute, Russian Academy of Sciences, Moscow, Russia*
[b] *Western Department of Pushkov Institute of Terrestrial Magnetism, Ionosphere, and Radio Wave Propagation, Russian Academy of Sciences, Kaliningrad, Russia*
[c] *A.M. Obukhov Institute of Atmospheric Physics, Russian Academy of Sciences, Moscow, Russia*
[d] *K.Tsiolkovsky State Museum of the History of Cosmonautics, Kaluga, Russia*
[e] *Saint-Petersburg State University, Saint-Petersburg, Russia*
[f] *Lomonosov Moscow State University, Moscow, Russia*
[g] *Silaeder School, Moscow, Russia*
[h] *Omsk educational institution "GDDYuT", Omsk, Russia*
[i] *Saint-Petersburg Department of Pushkov Institute of Terrestrial Magnetism, Ionosphere and Radio Wave Propagation of the Russian Academy of Sciences, Saint-Petersburg, Russia*
[j] *Unaffiliated persons*

*Corresponding author e-mail: ougolnikov@gmail.com

**Abstract**
During the summer season of 2025, noctilucent clouds (NLC) were observed at the latitudes 55-60°N from the late May until the late August. A distinct 5-day periodicity in their occurrence emerged following the summer solstice. Analysis of EOS Aura/MLS satellite data revealed that this effect was driven by a westward 5-day planetary wave, the amplitude of which was twice that of any previous northern summer since the start of the EOS Aura measurements in 2005. This study details the evolution of this exceptional planetary wave throughout the summer. Furthermore, NLC altitudes were determined via triangulation and colorimetry and were compared with MLS temperature profiles, enabling the determination of a mean positive phase lag for NLC occurrence relative to the temperature minimum.

## 1. Introduction

Noctilucent clouds (NLC) form in the summer upper mesosphere at mid- and polar latitudes, at the altitudes of approximately 80-85 km, where temperatures can drop to the ice freezing point of 145-150 K. The proximity of the temperature to this threshold makes NLC highly sensitive to variations in physical conditions, such as the passage of planetary waves (PW) and internal gravity waves (IGW). These waves have periods ranging from minutes to hours (for IGW) and can reach several days (for PW), with temperature amplitudes of several kelvins. IGW are known to directly modulate the visible structure of NLC (Ugolnikov, 2023a), as temperature fluctuations alter ice particle size and cloud brightness (Rusch et al., 2017; Gao et al., 2018).

Planetary waves, characterized by longer periods, govern the large-scale temperature distribution in the upper mesosphere and can induce periodic oscillations in NLC occurrence. Observed wave periods vary from 2 days (Azeem et al., 2001; Dalin et al., 2008, 2011) to 16 days (Espy et al., 1997; Luo et al., 2000), and 23 days (Pancheva & Mitchell, 2004). Longer 27-day oscillations in NLC occurrence (Robert et al., 2010) are linked to the solar rotation cycle.

The most prominent is (1,1) Rossby normal mode (Salby, 1981ab), with a period close to 5 days. The same periodicity was reported in visual NLC observations by Gadsden (1985), and a correlation between NLC occurrence and 5-day planetary wave activity was later established through long-term visual records (Sugiyama, 1998; Kirkwood & Stebel, 2003; Dalin et al., 2008, 2011). Although the amplitude of the 5-day wave in mesospheric temperature is strongest during the winter (Pancheva & Mitchell, 2004), it persists after the summer solstice – typically with an amplitude not exceeding 2 K – and increases towards the end of the summer (Liu et al., 2015). This relatively small and variable amplitude, combined with fluctuations in the wave period, has complicated the determination of the specific wave phase associated with the maximum probability of NLC occurrence.

Until the end of the 20th century, global data on mesospheric temperatures were insufficient to reconstruct the complete spatial-temporal structure of the 5-day wave. Kirkwood and Stebel (2003) attempted to extrapolate the wave phase from stratospheric data, but the resulting phase relationship between NLC occurrence and temperature remained uncertain.

The advent of satellite measurements in the early 21st century provided a solution. The 5-day wave was successfully detected in mesospheric temperature data from TIMED/SABER (Riggin et al., 2006) and Odin (Belova et al., 2008), as well as in polar mesospheric summer echoes analyses (Kirkwood et al., 2002; Iimura et al., 2015). Furthermore, satellite observations from SNOE and AIM-CIPS revealed a westward-propagating 5-day periodicity in the polar mesospheric cloud (PMC) field (Merkel et al., 2003, 2008, 2009). Analysis of NLC occurrence from SCIAMACHY data also showed 5-day variations, which were anti-correlated with temperature (von Savigny et al., 2007).

However, the phase relationship between the 5-day temperature wave and NLC/PMC occurrence remains an open question. Liu et al. (2015) identified a negative phase shift, implying that NLC are mostly observed several hours before the temperature minimum. This counterintuitive finding seems to require explanation by additional physical mechanisms. The issue could be clarified by analyzing a case of a strong, long-lived planetary wave with a stable period occurring in the summer mesosphere. The primary goal of this study is to conduct such an analysis for a notable event that took place in the northern summer of 2025.

## 2. NLC observations

Noctilucent clouds were observed by a network of all-sky and wide-angle cameras located at latitudes 54-58°N and longitudes 36-40°E. The number of operational observation sites in this region varied nightly and was up to seven in 2025. This dataset, described by Ugolnikov et al. (2025), enables the geometric determination of the altitude of individual cloud fragments. The technique (Ugolnikov, 2024) is based on a correlation analysis of cloud fields from different cameras, which are projected onto a surface corresponding to an *a priori* NLC altitude. Several altitude measurements were obtained using another two-camera system located near 60°N, 30°E at 80 km from each other.

When an extended cloud ensemble is captured by an all-sky RGB camera with known spectral characteristics for its optics and detectors, the altitude of the clouds can also be estimated by analyzing the dependence of color indices on the local solar zenith angle (Ugolnikov, 2023b). These color indices change as the cloud descends into the shadow cast by stratospheric ozone and, subsequently, the dense troposphere as the Sun approaches the limb of the Earth. The color analysis also allows for the determination of the effective radius of NLC particles.

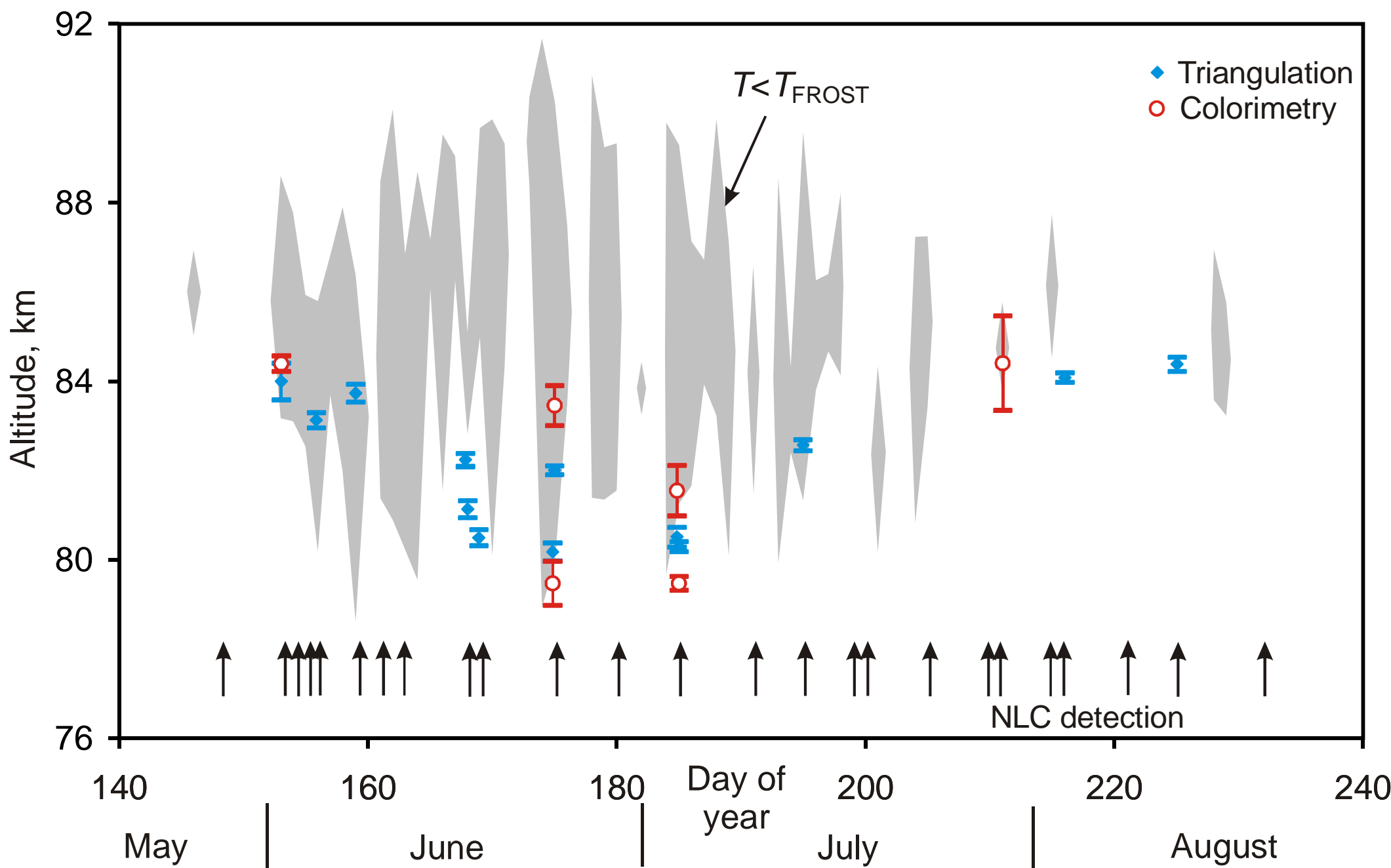

*Figure 1. The moments of NLC detection, mean triangulation and colorimetric altitudes (if measured) of the clouds during the summer 2025 fixed by the camera set 54-58°N, 36-40°E compared with nighttime supersaturated areas by MLS data. Note: altitudes for days of year 195.0 and 216.0 refer to the triangulation measurements at 60°N, 30°E.*

Strong noctilucent cloud activity was not anticipated for 2025 due to the solar cycle maximum, this inverse relationship is well-known (Gadsden, 1998; Kirkwood and Stebel, 2003; Romejko et al., 2003). However, the NLC season at the observation sites began surprisingly early. Following the first appearance in the morning of May 28, the clouds were observed almost nightly in early June. The timeline of NLC observations is presented in Figure 1 (where the Day of Year value of 1.0 corresponds to January 1, 00h UT). Arrows at the bottom of the figure indicate the nights during which NLC were detected by at least one camera. The figure also displays the results of successful triangulation or colorimetric altitude measurements if they were conducted in those nights. Two distinct altitudes for a single night correspond to separate analyses of the evening and morning twilight periods.

The early-season NLC observed in June were located at approximately 84 km, which is higher than the seasonal average. A similar elevation was noted for the late-season clouds, which were observed until the morning of August 20. This is a known effect, attributed to the temperature minimum in the summer mesosphere occurring near the solstice. Visible clouds are formed by particles that reach their maximum size near the bottom of the atmospheric layer where the temperature falls below the frost point. This supersaturated layer reaches 80 km and below around the solstice.

To illustrate this relationship, we utilized temperature and water vapor data from the EOS Aura/MLS satellite (NASA GES DISC, 2025). The spacecraft has a Sun-synchronous orbit (Schoeberl et al., 2006), providing one daytime and one nighttime measurement per day for any given atmospheric location. We interpolated the nighttime data for a longitude of 37°E and averaged it over latitudes 55-60°N. The resulting temperature was then compared to the frost point temperature, calculated from the $H_2O$ data using an empirical formula (Murphy & Koop, 2005). In Figure 1, atmospheric regions where the temperature is below the frost point are shaded in grey. It is evident that most NLC with measured altitudes are located near the bottom of these supersaturated regions.

In addition to their early and frequent appearance in June, the NLC exhibited another remarkable property: a pronounced 5-day periodicity in their occurrence, which is also visible in Figure 1. This periodicity began near the solstice and persisted for the remainder of the summer season until the early August. A corresponding periodicity is also evident in the temperature data. This pattern can be interpreted as the manifestation of a strong and stable planetary wave and will be used to analyze the phase relationship between NLC occurrence and temperature. This analysis, based on global EOS Aura/MLS data, will be presented in the following chapter.

## 3. Planetary wave period and amplitude by MLS data

To investigate the potential influence of a planetary wave in the mesosphere during the summer of 2025 and to compare it with previous years, we constructed a spectrum of zonal temperature variations. We used EOS Aura/MLS data at the 0.46 Pa pressure level (~83 km), which is closest to the typical NLC observation altitude, averaged over the latitude band of 55°N to 60°N. The temperature, $T$, is treated as a function of time, $t$, and longitude, $\lambda$. If we consider the value of $\tau$ as a possible period of the planetary wave, and the wavenumber is $k$, then we calculate the amplitude $A$ as follows:

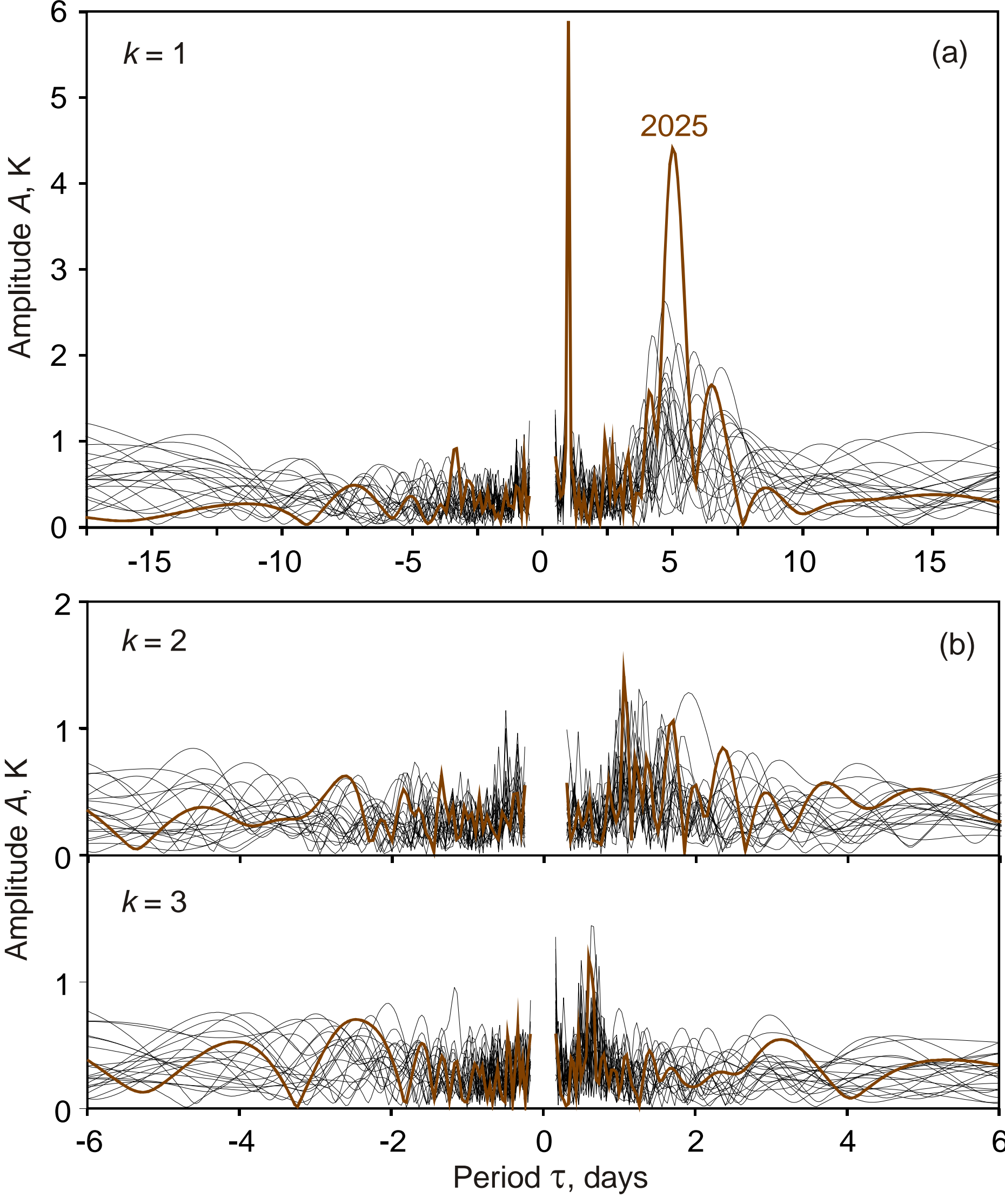

*Figure 2. The spectrum of the planetary wave during 30 days after the summer solstice for the wavenumber 1 (a), 2, and 3 (b), 55-60°N, 0.46 Pa in 2025 (bold) compared with 2005-2024 by EOS Aura/MLS data.*

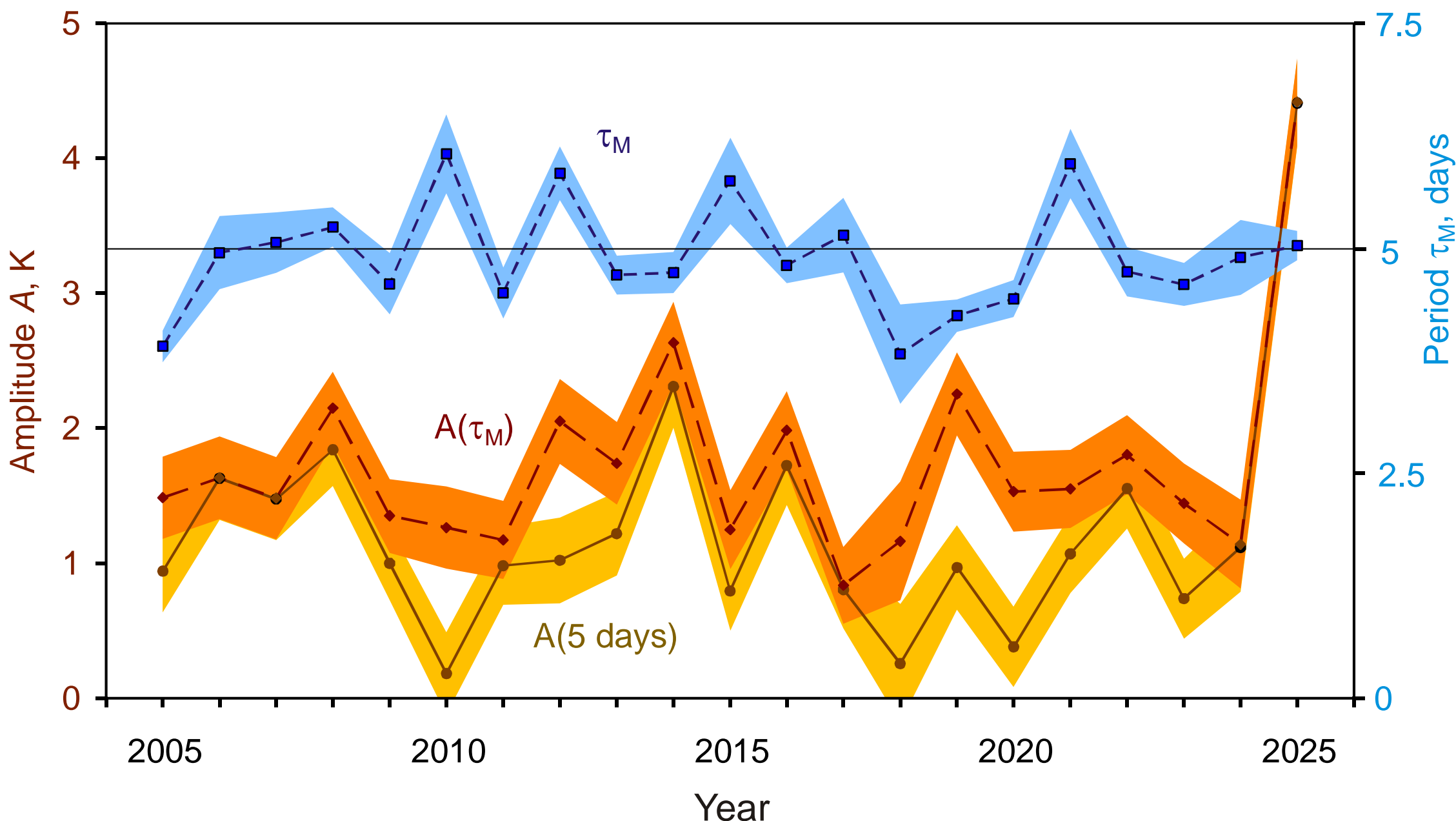

*Figure 3. The period $\tau_M$ corresponding to the maximal amplitude and the amplitudes for this period and for 5 days for the interval 30 days after solstice, 55-60°N, 0.46 Pa, 2005-2025, by EOS Aura/MLS data, error boxes are shown.*

$$X(k,\tau) = \frac{2}{N}\sum (T(\lambda,t) - T_0)\cdot\cos\left(k\cdot(\lambda + \frac{2\pi\cdot t}{\tau})\right);$$

$$Y(k,\tau) = \frac{2}{N}\sum (T(\lambda,t) - T_0)\cdot\sin\left(k\cdot(\lambda + \frac{2\pi\cdot t}{\tau})\right);$$

$$A(k,\tau) = \sqrt{X^2(k,\tau) + Y^2(k,\tau)}. \qquad (1)$$

Here the sums are computed over all MLS data points within a fixed time and latitude intervals, $T_0$ is the mean temperature over that interval, and $N$ is the total number of temperature measurements. This equation describes a westward-propagating planetary wave; an eastward wave can be investigated by considering $\tau<0$. If the temperature field was driven by a monochromatic, sinusoidal planetary wave, $A$ would correspond to its amplitude. An exception is the diurnal westward wave ($\tau$ =1 day), where the amplitude $A$ must be divided by a factor of 2 due to the Sun-synchronous orbit of EOS satellite and fixed solar times of measurements after the local noon and local midnight.

We analyzed the 30-day interval starting from the summer solstice, when the 5-day periodicity was noted in 2025 and when NLC occurrence is typically maximal. Figure 2a shows the spectra for the zonal wavenumber $k$ = 1 for all 21 summers of EOS Aura satellite operation from 2005 to 2025. Apart from the expected narrow maximum associated with diurnal variations ($\tau$ = 1 day), the most prominent feature is a strong peak at $\tau$ = 5 days in 2025. While enhancements at this period are visible in other years, the 2025 amplitude is approximately twice the highest value recorded in the previous two decades!

Amplitudes of the waves with higher wavenumbers (2 and 3) are shown in Figure 2b, they are basically below 1 K. The maximum value of the known 2-day oscillations for $k$ = 2 was fixed in 2018, when bright NLC were observed near the solstice in 2-day interval (Ugolnikov & Maslov, 2019), however, the amplitude was just about 1.3 K. This is possibly related to the short lifetime of 2-day oscillations, less than analyzed range of 30 days.

Unusual magnification of the 5-day wave is illustrated in Figure 3, which plots the 5-day wave amplitude, the maximum amplitude in the period range of 5.0 ± 1.2 days, and the period $\tau_M$ corresponding to this maximum amplitude for each year from 2005 to 2025. By varying the latitude and altitude, we build the maps of the 5-day wave amplitude for these years (Figure 4). The principal difference in the wave structure in 2025 compared to previous years demonstrates that the wave amplification was not a local phenomenon. Although the amplitude peaks near the mesopause over the Polar Circle, the highest amplitude at the altitude of noctilucent clouds is observed in the mid-latitude belt. This confirms the connection between the observed 5-day periodicity in NLC occurrence and the westward planetary wave, providing an opportunity to study the wave evolution and its role in NLC formation.

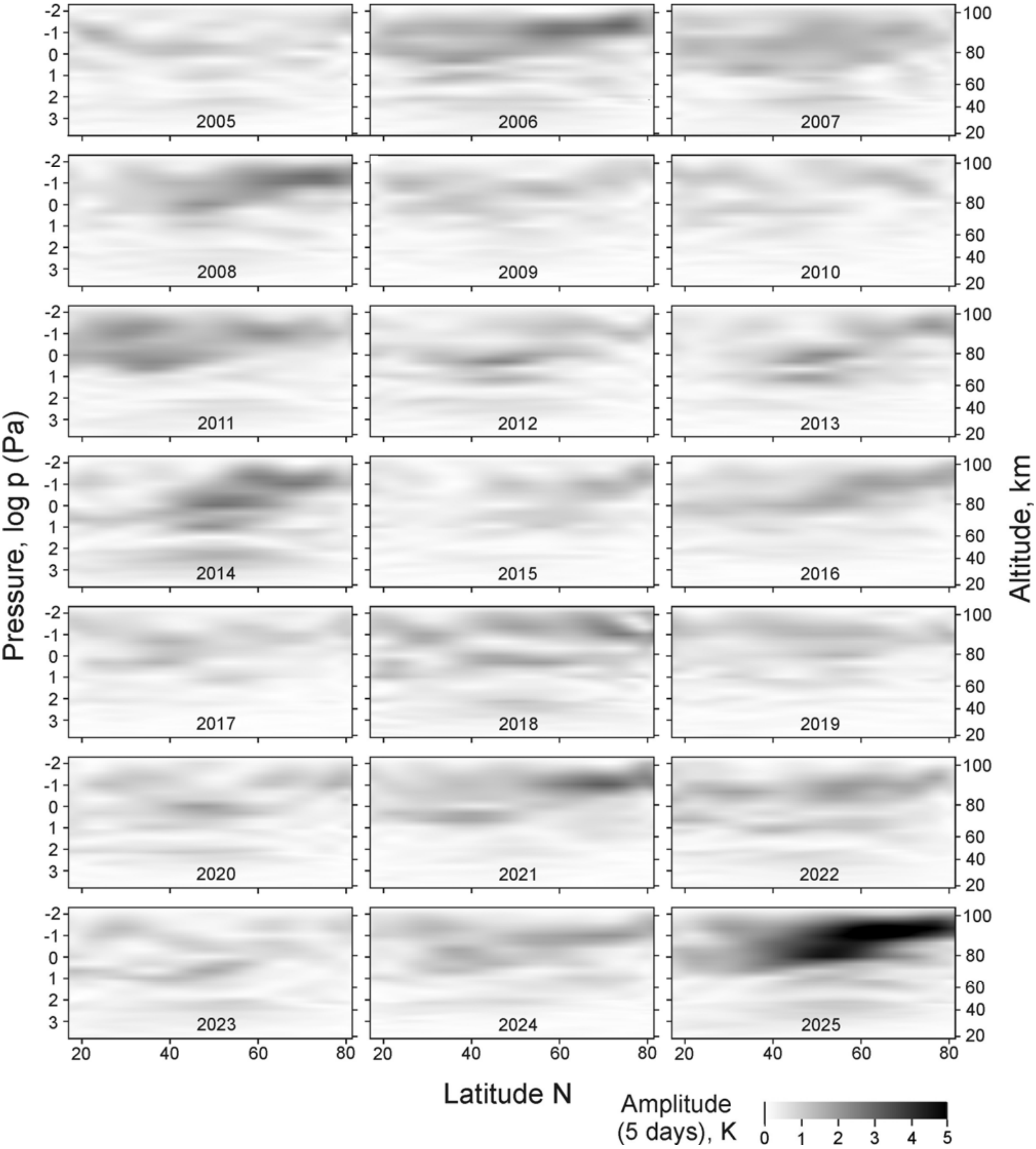

*Figure 4. Maps of the amplitude of the westward 5-day mode during the 30-day interval after the summer solstice by EOS Aura/MLS temperature measurements in 2005-2025.*

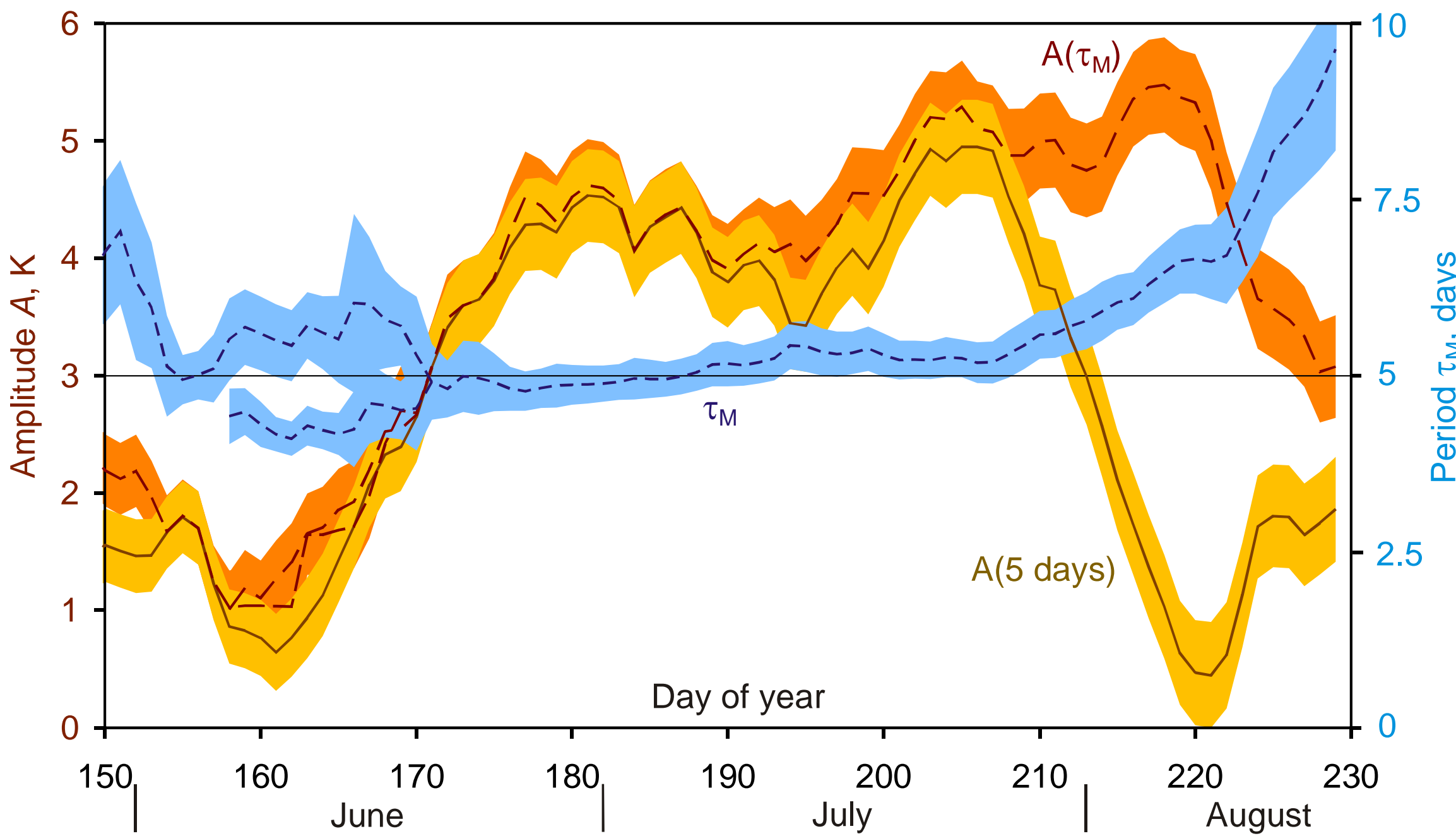

*Figure 5. The period corresponding to the maximal amplitude, $\tau_M$, and the amplitudes for this period and for 5 days during the summer of 2025, 55-60°N, 0.46 Pa, by EOS Aura/MLS data. For the time prior to day 171, the data for two periods $\tau_M$ corresponding to the different maxima of amplitude, are shown.*

To examine this during the summer of 2025, we computed the wave amplitude at 0.46 Pa and latitudes from 55°N to 60°N using a sliding 20-day window (±10 days from a central date) across the summer. Figure 5 shows the period $\tau_M$ with the maximum amplitude, along with the amplitude at this period and at the fixed period of 5 days. We see that the wave was not clearly expressed before the solstice: there were two similar maxima of spectrum $A(\tau)$ below and above 5 days, both shown in the figure. As noted earlier, NLC during this time were frequent but did not exhibit a clear periodicity.

Immediately at the solstice, these two maxima had merged into a single peak with a period close to 5 days and an amplitude increasing to 4 K. This is precisely when the periodic NLC occurrence began. The period of the strong planetary wave remained stable near 5 days until the end of July, when its amplitude peaked at approximately 5 K. In August, the wave amplitude began to decrease, while the period of maximum amplitude increased. As shown in Figure 1, noctilucent clouds were observed near the epochs of the temperature minima of this wave at the longitudes of the observation sites, enabling a detailed phase analysis to be conducted.

## 4. Phase analysis of NLC occurrence

As illustrated in Figure 5, the wave period remained remarkably stable for a 35-day interval starting from the solstice (encompassing seven full wave cycles, from day of year 172.0 to 207.0). This stable interval was selected for determining the phase relationship between the wave and NLC observations. Applying the procedure defined by Equation (1), we determined the period of maximum amplitude $\tau_0$ = 5.08 ± 0.15 days, with a mean amplitude $A_0$ = 4.2 ± 0.3 K. The wave mode with $k$ = 1 can be described by the equation:

$$T(\lambda,t) = T_C(t) - A_0 \cos\left(\lambda + 2\pi \frac{t-t_0}{\tau_M}\right) = T_C(t) - A_0 \cos\left(2\pi \frac{f}{\tau_M}\right) \quad (2)$$

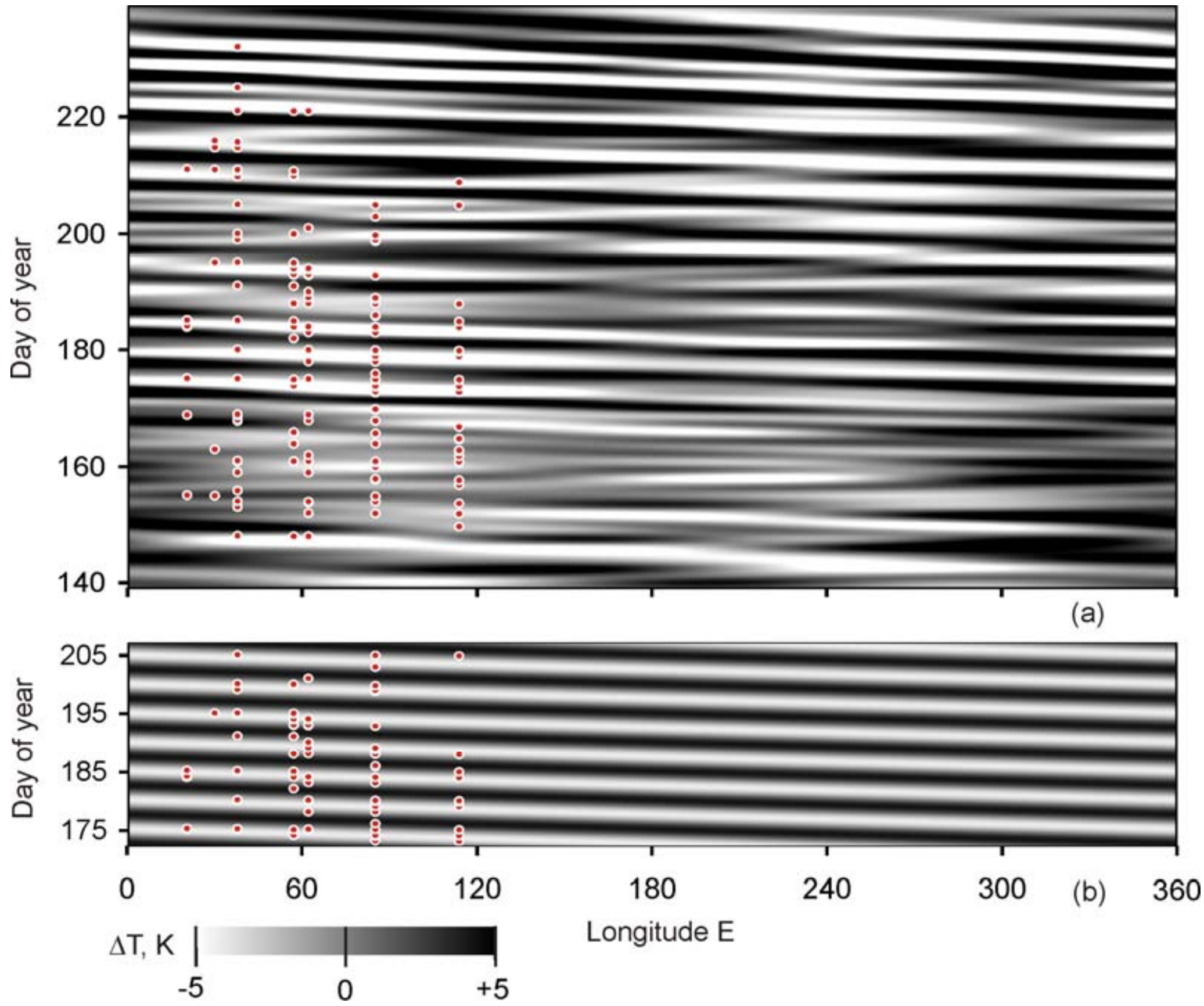

*Figure 6. (a) Temperature variations at 55-60°N, 0.46 Pa during the summer of 2025, the annual trend is removed. (b) The same for the monochromatic wave with the period 5.08 days. The moments of NLC detection are shown by the dots.*

Here, $T_C(t)$ represents the seasonal change of the mean temperature, modeled as a second-degree polynomial in time $t$. The parameter $t_0$ denotes the moment of the temperature minimum at the prime meridian (zero longitude), determined via a least-squares fitting procedure to be $t_0 = 179.92 \pm 0.06$ days (corresponding to June 28, 22:00 UT). The last equation is the definition of the wave phase $f$ as the time elapsed since the most recent (or until the following) temperature minimum at a fixed longitude $\lambda$:

$$f = (\lambda \tau_M / 2\pi) + (t - t_0) \qquad (3)$$

Since the period $\tau_M$ is close to an integer number of days and NLC are only visible at specific local solar times (during twilight), observations from a single location could cluster around similar phase values. To mitigate this potential sampling bias and to enrich the NLC occurrence dataset, we add the moments of NLC detection by five StarVisor.net project cameras at close latitude and longitudes from 20°E to 114°E.

Figure 6a shows the temperature variations at the 0.46 Pa level within the 55-60°N latitude band, with the seasonal trend $T_C(t)$ removed. The moments of NLC detection by all cameras during the summer period are also shown. Here and later, when NLC were observed during both the evening and morning twilight of the same night, the times are plotted separately. A visual inspection suggests that the westward 5-day wave with the wavenumber 1 is the primal mode driving the temperature. Most clouds were observed slightly after the temperature minimum. For an exact analysis, we focus on the interval from day 172.0 to 207.0 and the wave mode with the period $\tau_M$, as shown in Figure 6b (with the polynomial trend $T_C(t)$ removed). The phase $f$ was calculated for each NLC detection event within this interval and is presented in the circular histogram in Figure 7. The points are plotted with different radii corresponding to their observation longitudes for clarity.

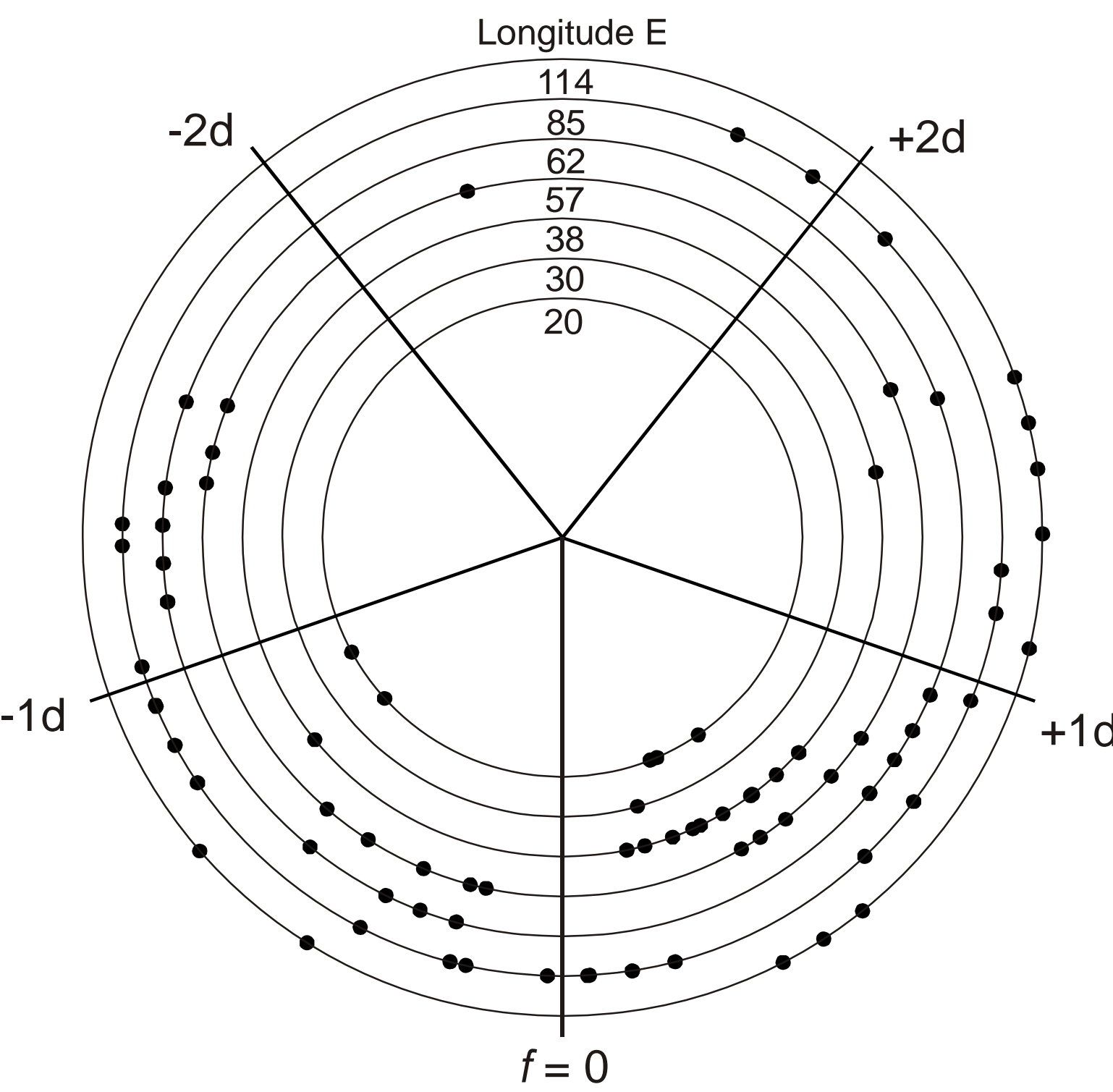

*Figure 7. The phase histogram of NLC occurrence for the interval 35 days after the summer solstice 2025, for the period 5.08 days. Different circles correspond to the different longitudes of the observations listed in the top of the figure.*

The phase distribution reveals that NLC occur predominantly near the zero phase of the wave ($f = 0$), but with a notable asymmetry; detections with positive phases (after the temperature minimum) are more frequent. This indicates that the peak in NLC occurrence probability lags slightly behind the temperature minimum. The simplest, though not most robust, quantitative estimate of this lag is the arithmetic mean of the phases, which is +0.10 days. A more rigorous method involves calculating the average direction of the unit vectors in the circular histogram (Figure 7), which yields a mean NLC occurrence phase of $+0.14 \pm 0.09$ days or $+3.4 \pm 2.2$ hours. This measured positive phase lag is seen in the 2025 data. Following the summer solstice, the temperature minimum of the 5-day planetary wave crossed the longitude of the central camera set (37°E) around the local noon on June 23, 28, July 3, etc, while bright and expansive noctilucent clouds were observed in the nights immediately following (see Figure 1). This illustrates the positive phase lag identified in our analysis.

## 5. Discussion and conclusion

The high noctilucent cloud activity observed during the summer of 2025 was unexpected due to the solar 11-year cycle maximum (the period typically associated with suppressed NLC occurrence). This anomaly also extended to the amplitude of the 5-day planetary wave, which similarly decreased after the maximum of solar activity (Gadsden, 1998). Despite these predictions, NLC were frequently observed at latitudes of 55-60°N over an extended season, from the late May until the second half of August. For a number of these events, cloud altitudes were successfully measured via triangulation and/or RGB photometry. The change of observed elevation of NLC altitudes during the summer is consistent with known mesospheric temperature dynamics and must be accounted for when investigating potential long-term trends in NLC altitude, possibly driven by atmospheric shrinking effects (Lübken et al., 2018).

The final altitude measurement in 2025 was recorded on the morning of August 13. A notable increase in NLC activity during the second decade of August, potentially linked to the Perseid meteor shower, has been previously documented in both high-latitude ground observations (Ugolnikov et al., 2016, 2017) and SCIAMACHY satellite data (Robert et al., 2009).

Analysis of EOS Aura/MLS temperature data confirms that the 5-day planetary wave in the summer of 2025 was the most powerful of the last two decades. Interestingly, the previous maximum summer amplitude for this wave occurred 11 years prior, in 2014, yet it was a half the strength observed in 2025 (see Figures 2-4). While this could be a coincidental effect, it underscores the need for further investigation into the complex relationship between planetary wave activity and the solar cycle. In 2025, the wave amplified abruptly near the summer solstice and persisted throughout the season, completing over ten revolutions around the North Pole. Its temperature amplitude at NLC altitudes peaked at 5 K in July.

A key feature of the 2025 event was the remarkable stability of the wave period, which remained close to 5 days for the first 35 days post-solstice. This stability provided a unique opportunity to conduct a robust phase analysis of NLC occurrence using a distributed network of cameras across a wide longitudinal range, the question had remained challenging in previous studies. The analysis revealed a positive mean phase lag, indicating that the maximum probability of NLC occurrence follows the temperature minimum of the wave by approximately 3 hours. Considering the mean effective particle radius of 65 nm derived from RGB photometry in 2025 and a typical water vapor mixing ratio of 5 ppm from MLS data, this phase lag corresponds to roughly half the characteristic particle growth time (approximately 6 hours, Gadsden and Schröder, 1989). In other words, the NLC occurrence peak aligns with the time when the midpoint of the particle growth period coincides with the most favorable (coldest) conditions.

The drivers behind the unusual and unpredicted dynamics of the summer mesosphere in 2025 remain an open question. With the conclusion of the AIM satellite mission, future investigations into NLC-wave interactions will rely not only on global satellite data for mesospheric temperature and polar mesospheric clouds but also on expanded ground-based observations from diverse longitudes. Such a collaborative, multi-platform approach is essential to enhance our data on NLC altitudes and microphysics, ultimately enabling a clearer understanding of their long-term trends and responses to a changing atmosphere.

**Acknowledgments**

Authors would like to thank Natalia Andrievskaya, Olga Borchevkina, Dmitry Aleshin, Peter Dalin, Alexander Ilykhin, Vyacheslav Ignatiev, Dmitry Kobets, Vadim Lymar, Denis Maximov, Anatoly Morozov, Dmitry Moskin, Natalia Nikishkina, Roman Semenyk, Yuri Stepanychev, Alexandr Timchenko, Alexej Tkachev, Vitaly Tuh, and Mikhail Zavyalov for their help in noctilucent clouds multi-site survey.